\documentclass[lettersize,journal]{IEEEtran}
\usepackage{amsmath,amsfonts}
\usepackage{algorithmic}
\usepackage{algorithm}
\usepackage{array}
\usepackage{amssymb} 
\usepackage{pifont}  
\usepackage[caption=false,font=normalsize,labelfont=sf,textfont=sf]{subfig}
\usepackage{textcomp}
\usepackage{stfloats}
\usepackage{url}
\usepackage{hyperref}
\usepackage{verbatim}
\usepackage{graphicx}
\usepackage{cite}
\usepackage{multirow,enumitem}
\usepackage[mathscr]{eucal}
\usepackage{booktabs}
\usepackage{footmisc}
\usepackage{xcolor}
\usepackage{bbding}

 \usepackage[thinc]{esdiff}
 \usepackage{xcolor}
 \usepackage{subfig}
 \usepackage{bbding}
\newcommand{\be}{\begin{eqnarray}}
\newcommand{\ee}{\end{eqnarray}}

\usepackage{comment}
\usepackage{url}
\usepackage{colortbl}
\usepackage{algorithm}
\usepackage{algorithmic}
\usepackage{amsmath}
\usepackage{MnSymbol,bbding,pifont}
\def\x{{\mathbf x}} 
\def\e{{\mathbf e}}
\def\s{{\mathbf s}}

\newcommand{\method}{\textcolor{black}{S2S-ZEST}}

\begin{document}

\title{Textless and Non-Parallel \textcolor{black}{Speech-to-Speech} Emotion Style Transfer}

\author{Soumya Dutta,~\IEEEmembership{Student Member,~IEEE,} 
        Avni Jain and 
        Sriram Ganapathy,~\IEEEmembership{Senior Member,~IEEE}
\thanks{S. Dutta and S. Ganapathy are with the learning and extraction and acoustic patterns (LEAP) laboratory, Electrical Engineering, Indian Institute of Science, Bangalore, India, 560012. Avni Jain is with the Department of Computer Science, Birla Institute of Technology and Science (BITS Pilani) Goa campus. This work was done during her internship at LEAP lab. This work was performed with grants received as part of the prime ministers research fellowship (PMRF), Ministry of Education, India as well as grants from British Telecom. }
\thanks{Manuscript received Oct xx, 2025.}}

\markboth{Journal of \LaTeX\ Class Files,~Vol.~14, No.~8, Oct~2025}%
{Shell \MakeLowercase{\textit{et al.}}: A Sample Article Using IEEEtran.cls for IEEE Journals}


\maketitle

\begin{abstract}
Given a pair of source and reference speech recordings, \textcolor{black}{speech-to-speech (S2S)} emotion style transfer  involves the generation of an output speech that mimics  the emotion characteristics of the reference   while preserving the content and speaker attributes of the source.
\textcolor{black}{In this paper, we propose a speech-to-speech zero-shot emotion style transfer framework, termed as} \textcolor{black}{S2S} {\textbf{Z}ero-shot  \textbf{E}motion \textbf{S}tyle \textbf{T}ransfer (\textbf{\method})},  that enables the transfer of emotional attributes from the reference to the source while retaining the speaker identity and speech content.
The \textcolor{black}{S2S-ZEST} framework consists of an analysis-synthesis pipeline in which the analysis module \textcolor{black}{extracts semantic tokens, speaker representations, and emotion embeddings from speech.} 
 Using these representations, a pitch contour estimator and a duration predictor are learned. Further, a synthesis module is  designed to generate speech based on the input representations and the derived factors. \textcolor{black}{The analysis–synthesis pipeline is trained using an auto-encoding objective to enable efficient resynthesis during inference.}
For \textcolor{black}{S2S} emotion style transfer, the emotion embedding extracted from the reference speech  along with the rest of the representations from the source speech are used in the synthesis module to generate the style translated speech. 
In our experiments, we evaluate the converted speech on content/speaker preservation (w.r.t. source) as well as on the effectiveness of the emotion style transfer (w.r.t. reference). \textcolor{black}{The proposed framework demonstrates improved emotion style transfer performance over prior methods in a textless and non-parallel setting.} We also illustrate the application of the proposed work for data augmentation in emotion recognition tasks.
\end{abstract}


\section{Introduction}\label{sec:intro}

\IEEEPARstart{E}{motion} Style Transfer (EST) aims to modify the emotional expression of a speech signal while preserving key attributes such as speaker identity and linguistic content~\cite{zhou2022emotional}. This technology has broad applications, particularly in human-computer interaction, where it enables machines to not only recognize but also respond to human emotions more naturally~\cite{pittermann2010handling,crumpton2016survey, rosenberg2021prosodic}. Despite recent advancements in speech synthesis, EST remains challenging due to the complexity of modeling human emotions and the difficulty of accurately transferring them in generation tasks~\cite{plutchik1991emotions}.\par

Early approaches for EST focused on modifying the spectral envelope and pitch contour (F0 contour) of neutral speech to generate emotional speech. Inanoglu et al.\cite{inanoglu2009data} employed a hidden Markov model (HMM) for F0 contour conversion and a Gaussian mixture model (GMM) for spectral transformation. Luo et al.\cite{luo2016emotional} utilized deep neural networks to convert mel-frequency cepstral coefficients (MFCC) and the F0 contour, while Lorenzo-Trueba et al.\cite{lorenzo2018investigating} explored deep learning architectures for emotional speech synthesis. A major limitation of these approaches is their reliance on parallel emotional speech datasets, where the same speaker articulates the same content in multiple emotions. Recording such data at scale is both expensive and challenging. \textcolor{black}{To overcome this limitation, a large body of prior work in voice conversion and expressive speech synthesis has explored factorized representations of speech, in which content, speaker identity, and speaking style are modeled separately~\cite{polyak2021speech,kreuk2021textless,maimon2022speaking,oh2024durflex,yao2025stablevc}. Inspired by this established paradigm, emotion style transfer can be formulated as a mixing problem, where the content and speaker characteristics of a source utterance are combined with the emotional style of a reference utterance. A model's  ability to perform such mixing without explicit supervised training on parallel data is referred to as \textit{zero-shot emotion style transfer (ZEST)}}.\par

Building on this motivation, we propose a speech-to-speech emotion style transfer (EST) framework that enables zero-shot transfer of emotional speaking style from a reference to a source utterance. \textcolor{black}{Prior speech style transfer and voice conversion methods~\cite{oh2024durflex, yao2025stablevc, maimon2022speaking} often exhibit entanglement between speaker identity and expressive style in practice. As a result, these approaches often fail to transfer the emotion style to unseen speakers. A recent effort by Zhang et al.~\cite{zhang2025vevo} has attempted to address these challenges through improved disentanglement and control mechanisms. However, this approach relies on large-scale training data.} 

Our early effort attempted emotion style transfer through a factored representation learning setting \cite{dutta2024zero}. In this paper, we extend our efforts in multiple ways. Our approach leverages two modules - 
\begin{itemize}
    \item An analysis module that encodes the input speech into four key components: semantic tokens, speaker representations, pitch features, and emotion embeddings.
    \item A synthesis module, based on BigVGAN~\cite{leebigvgan}, that reconstructs the speech given the tokens, speaker embeddings, emotion embeddings and pitch features, which is fully trained using the reconstruction (auto-encoding) loss. 
\end{itemize}

The key contributions beyond the prior work~\cite{dutta2024zero} are:

\begin{itemize}
\item We propose the tokenization of soft-HuBERT~\cite{van2022comparison} features for speech content representation and employ BigV-GAN as a unit-to-speech vocoder, replacing the conventional HuBERT \cite{hsu2021hubert} and HiFi-GAN~\cite{kong2020hifi} models.
\item We use a duration estimator that allows the emotion and speaker embeddings to modify duration of the semantic tokens during inference. 
We further develop a pitch reconstruction technique that operates on duration-modified tokens, along with speaker and emotion information, to enhance the naturalness of the synthesized speech.
\item We conduct extensive subjective and objective evaluations, demonstrating that \method{} outperforms baseline methods in emotion style transfer.
\item We showcase the application of the \method{}  as a data augmentation tool for speech emotion recognition (SER). 
\end{itemize}
\section{Related work}\label{sec:related}

\subsection{Emotional Style Transfer for Text-to-Speech Synthesis}
One of the earliest works, by Wang et al.\cite{wang2018style}, introduced ``style tokens'' to learn the speaking style of a reference speech signal. Subsequent methods by Min et al.\cite{min2021meta} and Lee et al.\cite{lee21h_interspeech} further refined style transfer techniques, though they primarily operated at the sentence level. Generspeech\cite{huang2022generspeech} addressed this limitation by incorporating more localized style cues from reference speech. In contrast to these works, our work on \method{} operates in a completely textless setting in both training and style transfer.

\subsection{Synthesis of Speech from Discrete Tokens}
Speech synthesis from discrete tokens was explored with vocoder models, ranging from classical methods (Griffin-Lim~\cite{griffin1984signal}, WaveNet~\cite{vandenoord16_ssw}) to GAN-based models (HiFi-GAN~\cite{kong2020hifi}, BigVGAN~\cite{leebigvgan}) and diffusion-based approaches (DiffWave~\cite{kong2020diffwave}). Polyak et al.\cite{polyak2021speech} introduced a speech synthesis approach that bypasses mel-spectrogram reconstruction by encoding speech content using HuBERT\cite{hsu2021hubert}. These representations were concatenated with pitch and speaker representations to train a HiFi-GAN vocoder for speech synthesis, a method adopted in multiple studies~\cite{lee2021direct,lee2021textless,maimon2022speaking,kreuk2021textless}. In contrast, our work explores a purely zero-shot setting for the style transfer task  using components like duration, pitch, and emotion embeddings. 

\subsection{Style Transfer in Speech Synthesis}

Most prior works in speech style transfer modify both speaker identity and speaking style, and therefore fall under the ambit of voice conversion (VC). Notable examples include DISSC~\cite{maimon2022speaking}, PromptVC~\cite{yao2024promptvc}, StableVC~\cite{yao2025stablevc}, as well as methods that incorporate emotion cues such as Kreuk et al.~\cite{kreuk2021textless} and DurFlex-EVC~\cite{oh2024durflex}. These approaches   assume that speaking style is inseparable from speaker identity, leading to entangled representations that limit their ability to generalize to unseen speakers.

A recent work~\cite{zhang2025vevo} addresses this limitation by transferring emotional speaking style from a reference to a source utterance, termed \textit{VEVO}. In contrast, our model does not require any transcribed speech during training and is entirely textless.

\section{Proposed Approach}\label{sec:method}

\begin{figure}[t!]
    \centering
\includegraphics[width=0.47\textwidth,trim={0cm 2cm 3.5cm 2cm},clip]{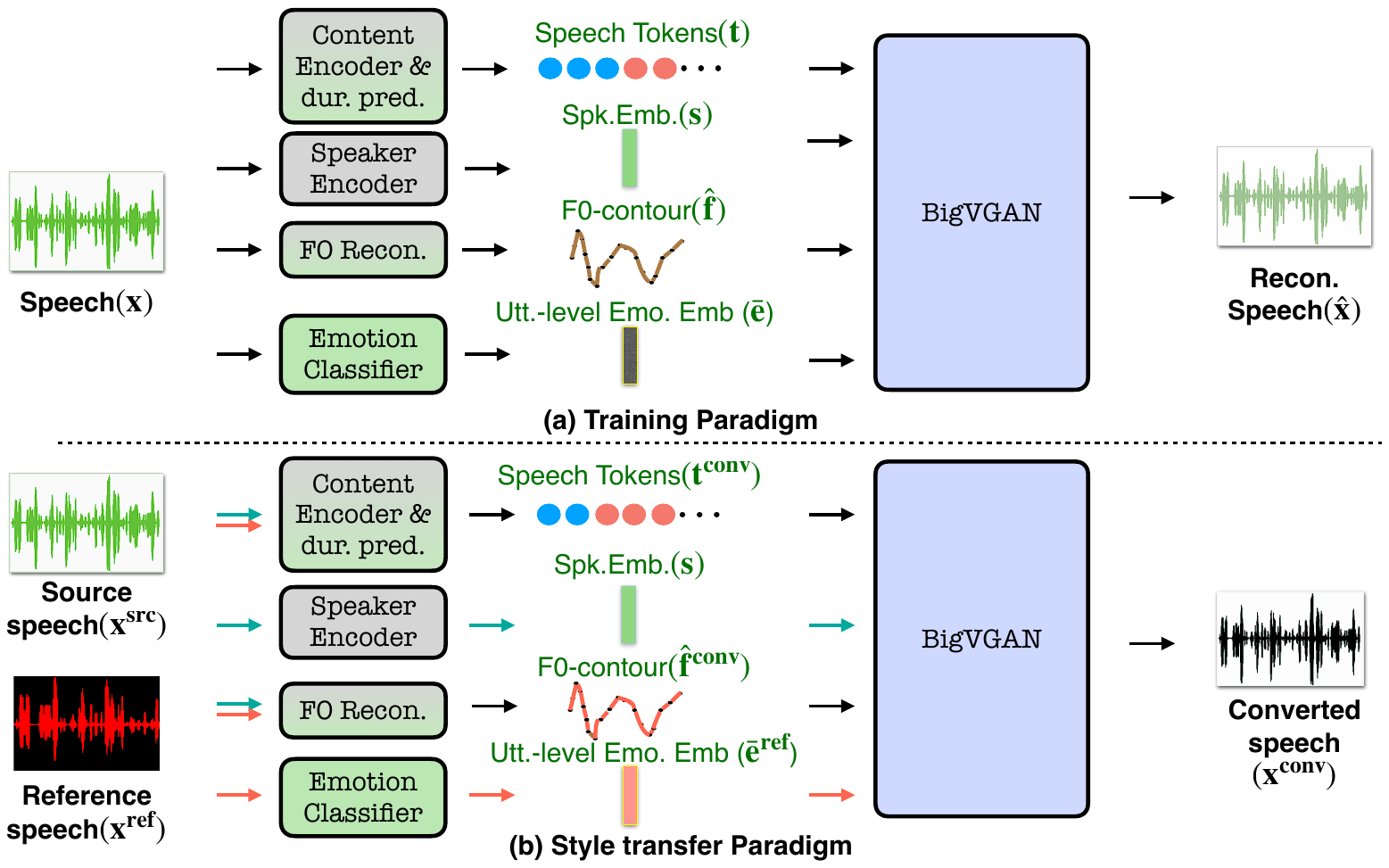}

    \caption{Overview of (a) \method{} training  and (b) style transfer paradigm. Emotion style factors are colored differently from the rest. During style transfer, the source speech tokens are passed through the content and speaker encoder, while the duration predictor, F0 reconstruction module and emotion classifier modules receive their input from the reference speech.}
    \label{fig:overview}

\end{figure}
\subsection{Big picture}

Let an input speech recording be denoted as $\x = \{x_1,...,x_N\}$, where $N$ denotes the number of samples in the recording. Following a windowing operation with $N_w$ samples in each frame, we obtain $T$ frames (for non-overlapping frames, $T=N/N_w$). 
The windowed speech signal is   converted to a sequence of tokens using the content encoder (detailed in Sec.~\ref{sec:con}). 
The tokenizer generates $\{ t_1,... t_T \}$ discrete tokens, where $t_i \in \{1...V\}, $ is the cluster index and $V$ is the total vocabulary size of the tokenizer. 
The speaker information is captured by an utterance-level embedding $\s$, extracted using a pre-trained speaker encoder (Sec.~\ref{sec:spkr}).  
 The emotion information in the signal is extracted with an emotion classifier, which provides frame-level   embeddings ($\mathbf{E} = \{\e_1, \e _2, .. \e _T\}$) and an utterance-level pooled output ($\mathbf{\bar {e}}$).  The emotion embedding extraction is  detailed in Sec.~\ref{sec:sace}.
 
Using the basic components (token sequence, speaker and emotion embeddings), we train,
 \begin{itemize}
     \item Pitch contour reconstruction module, which reconstructs the frame-level F0 sequence, $\mathbf{\hat{f}} = \{f_1,...f_T\}$, of $\mathbf{x}$, described in Sec.~\ref{sec:pitchpred}. 
     \item Token duration predictor, which generates the duration $\mathbf{\hat{d}} = d_1,...,d_{T^{'}}$ of the de-duplicated tokens $t_1,..t_{T^{'}}$. Here, $T^{'}$ denotes the number of tokens after de-duplication\footnote{E.g. if tokens are $\{1,1,1,41,41,1,1,5,5,5,5,5\}$ the de-duplicated token sequence is $\{1,41,1,5\}$ with the durations being $\{3,2,2,5\}$.}. This is  detailed in Sec.~\ref{sec:dur}. 
 \end{itemize}
Thus, the encoder analysis pipeline extracts five components from the speech signal: i) de-duplicated tokens, ii) token durations, iii) speaker embedding, iv) emotion embedding, and v) pitch contour. With these components as input, we train a BigVGAN model for reconstructing the speech signal using the auto-encoding loss (Sec.~\ref{sec:synth}). A brief overview of the training procedure for \method{} is shown in Fig.~\ref{fig:overview} (a). 

 During style transfer (depicted in Fig.~\ref{fig:overview} (b)),
  the source speech tokens  are generated and de-duplicated to obtain unique tokens, $t_1,..,t_{T^{'}}$. 
 The duration predictor is fed with source speech content and reference speech style information, and this generates the token durations,  $\mathbf{\hat{d}}^{conv}$. The speaker embedding, $\s$,  is extracted from the source speech, while the emotion embedding, $\mathbf{E}^{ref}$, is extracted from the reference speech. The F0 contour, $\mathbf{\hat{f}}^{conv}$, is constructed using the source content and speaker along with the reference emotion embeddings. These representations are input to the synthesizer to generate the style-converted speech. The emotion style-transfer process is elaborated in Sec.~\ref{sec:emoconv}.

\subsection{Content Encoder}
\label{sec:con}

In order to encode speech content that is devoid of speaking-style or speaker information, previous work by Polyak~\cite{polyak2021speech} illustrated the benefits of discretization of self-supervised speech embeddings. 
 Separately, Niekerk et al.\cite{van2022comparison}  refined a HuBERT-base model to predict frame-level soft embeddings with a continued pre-training of the HuBERT back-bone.  
However, the continuous soft-HuBERT features make it difficult to control the speaking rate.
We use the pre-trained soft-HuBERT embeddings  \cite{van2022comparison} and train a k-means clustering model to generate discrete tokens of speech. 
This preserves the intelligibility benefits of soft-HuBERT while allowing control over speech duration during synthesis. 
Denoting the output from the content encoder as $\mathbf{t}=\{t_1, t_2, \dots,t_T\}$, we have  $\mathbf{t} = \texttt{k-means}(\mathbf{H})$, where the soft-HuBERT model embeddings are denoted as, $\mathbf{H} = \{\mathbf{h}_1,...,\mathbf{h}_T\}$, i.e., $\mathbf{H} = \texttt{soft-HuBERT}(\x)$
\begin{figure}[t!]
    \centering
\includegraphics[width=0.45\textwidth,trim={3cm 11cm 6cm 3cm},clip]{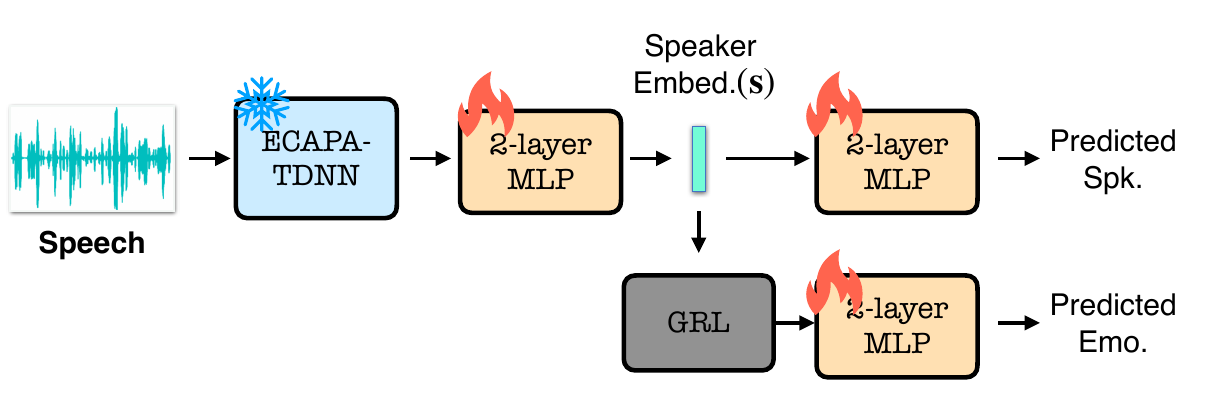}
    \caption{The model for extracting the speaker embedding. GRL stands for Gradient Reversal Layer. The blue block is kept frozen during training. }
    \label{fig:ease_model}
\end{figure}
\subsection{Speaker Encoder}
\label{sec:spkr}

To extract speaker embeddings, we employ a pre-trained speaker verification model. Specifically, we use ECAPA-TDNN~\cite{desplanques20_interspeech}, which generates x-vector embeddings by applying utterance-level pooling over frame-level representations.
Prior studies~\cite{pappagari2020x, shaheen23_interspeech} indicate that x-vectors also encode emotion information, which can interfere with style transfer. To mitigate this, inspired by Li et al.\cite{li2022cross}, we introduce two fully connected layers to the x-vector model and apply an emotion adversarial loss \cite{ganin2016domain}. The speaker embeddings in \method{} are trained using the following loss function:
\begin{equation}\label{eq:lossspkr}
    \mathrm{L}_{tot-spk} = \mathrm{L}_{ce}^{spk} - \lambda_{adv}^{emo} \mathrm{L}_{ce}^{emo}
\end{equation}
where $\mathrm{L}_{ce}^{spk}$ is the  speaker classification loss, and $\mathrm{L}_{ce}^{emo}$ is the emotion classification loss.  This is depicted in Fig.~\ref{fig:ease_model}. 

\subsection{Emotion Style Factors}
\label{sec:emo_pred}

\noindent \subsubsection{\textbf{Emotion classifier}}
\label{sec:sace}
We use an emotion classifier with the pre-trained HuBERT-base model, which comprises a convolutional feature extractor followed by $12$ transformer layers. 
For emotion classification, we fine-tune the transformer layers. 
The emotion embedding is the output of the last transformer layer, denoted as $\mathbf{E} = \texttt{Emo-embed}(\mathbf{x})$,
where $\mathbf{E} = {\{\e_1,...\e_T\}}, \e_i \in \mathrm{R}^{D}$,  with $D=768$. The frame-level embedding matrix $\mathbf{E}$ is average pooled over the temporal dimension to get the utterance-level emotion embedding $\mathbf{\bar{e}}$.  
 A softmax classification head is trained on $\mathbf{\bar{e}}$ for predicting the emotion class. We also require the  emotion embeddings to  be disentangled from the speaker attributes. To achieve this, we employ a speaker adversarial loss (similar to the speaker embedding extractor). The loss function is formulated as:
\begin{equation}\label{eq:lossemo}
    \mathrm{L}_{tot-emo} = \mathrm{L}_{ce}^{emo} - \lambda_{adv}^{spk} \mathrm{L}_{ce}^{spk}
\end{equation}
where $\mathrm{L}_{ce}^{emo}$ and $\mathrm{L}_{ce}^{spk}$ are the cross-entropy loss  functions for emotion and speaker classification,  respectively.
\begin{figure}[t!]
    \centering
\includegraphics[width=0.47\textwidth,trim={0cm 4cm 1cm 2cm},clip]{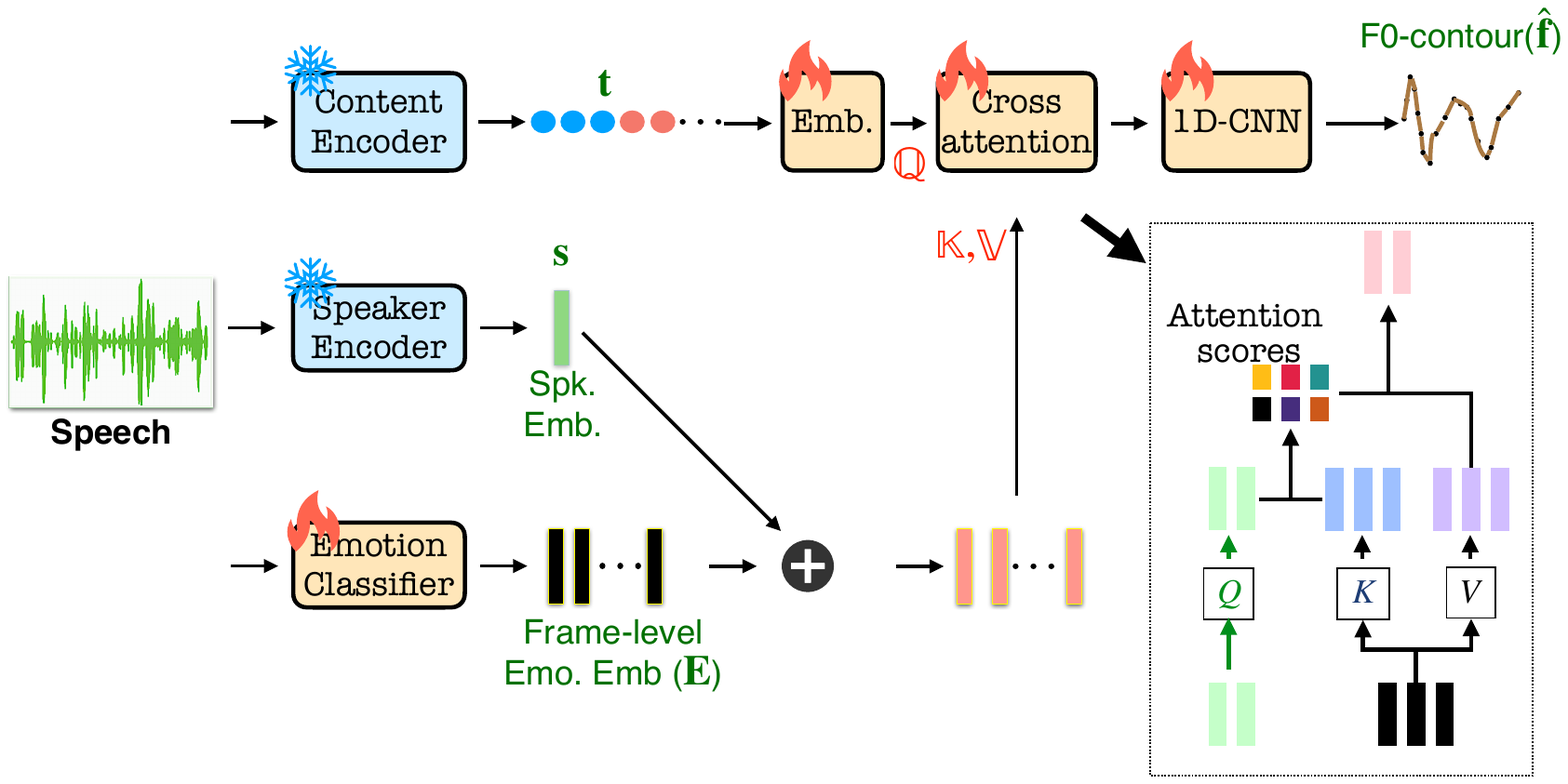}

     \caption{Pitch contour reconstruction module - Speaker embedding ($\mathbf{s}$) is added with frame-level emotion embeddings $\mathbf{E}$ and forms the key-value sequence while the source speech token embeddings ($\mathbf{C}$) form the query sequence. The frame-level outputs from the cross-attention block are passed through a position-wise feedforward network using 1D-CNNs to reconstruct the pitch contour ($\mathbf{\hat{f}}$). The cross-attention block architecture is also expanded for reference.}
    \label{fig:f0pred}

\end{figure}

\subsubsection{\textbf{Pitch Contour Reconstruction}}
\label{sec:pitchpred}
We propose a factored approach for pitch contour reconstruction, influenced by content, speaker, and emotion information. 
A learnable embedding layer provides the content representation $\mathbf{C}$ from the token sequence $\mathbf{t}$. 
As local pitch variations contribute to the emotional expressiveness~\cite{erro2009emotion,chen2024vesper}, we utilize the frame-level emotion embeddings $\mathbf{E}$ along with content representations $\mathbf{C}$ and speaker embedding $\mathbf{s}$.   
The pitch reconstruction module based on cross-attention and a $1$-D CNN network, is formulated as:
\begin{equation}\label{eq:f0}
    \mathbf{\hat{f}} = \texttt{1D-CNN}(\texttt{Attn}(\mathbf{C},~~ \mathbf{s}+\mathbf{E},~~\mathbf{s}+\mathbf{E} ) )
\end{equation}
In the above equation, the query representations $\mathbf{Q}$ come from the  content embeddings, while the speaker embedding $\mathbf{s}$ and the frame-level emotion embeddings $\mathbf{E}$ are added together to form the key ($\mathbf{K}$) and value ($\mathbf{V}$) representations. \textcolor{black}{In the proposed  formulation, content embeddings are used as queries in the cross-attention model, while emotion and speaker embeddings are used as keys and values, respectively.
    This design allows the model to condition the pitch generation on  prosodic patterns of the target emotion and the source speaker identity, while anchoring the temporal alignment to the spoken source content.}
The target pitch contour $\mathbf{f}=\{f_1, f_2,f_3,\dots,f_T\}$ is extracted using the YAAPT algorithm~\cite{kasi2002yet}, and the loss for pitch reconstruction module is defined as:
\begin{equation}\label{eq:lossf0}
    \mathrm{L}^{f0} = | |\mathbf{f}-\mathbf{\hat{f}}||_{L_1}
\end{equation}
\textcolor{black}{Unvoiced frames identified by the YAAPT algorithm are assigned a value of zero and are included in the loss computation.} This gives us a way to derive pitch contours based on content and speaker from one speech signal and emotion from another speech signal.  
The block diagram for pitch reconstruction is shown in Fig.~\ref{fig:f0pred}.

\subsubsection{\textbf{Duration Prediction}}
\label{sec:dur}
The discretization of self-supervised learning (SSL) features $\mathbf{H}$,   often results in repeated tokens~\cite{maimon2022speaking, lee2021direct, lee2021textless, kreuk2021textless}. To address this, we de-duplicate the tokens from the content encoder, $\mathbf{t}^{'} = \texttt{de-dup}(\mathbf{t})$.
During style transfer, we require the duration (speaking rate) of each de-duplicated source token conditioned on the speaker identity and the reference speech emotion~\cite{lee2021direct, maimon2022speaking}.  
To achieve this objective, we train the duration predictor module $\mathtt{D_{pred}}$ as:
\begin{equation}\label{eq:dur}
    \mathbf{\hat{d}} = \mathtt{D_{pred}}(\mathbf{\mathbf{t}^{'}}, \mathbf{s}, \mathbf{\bar{e}}) 
\end{equation}
The $\mathtt{D_{pred}}$ module consists of a learnable token-to-embedding layer (for inputs $\mathbf{t}^{'}$). The token embeddings are concatenated with utterance-level speaker embedding $\mathbf{s}$ and emotion embedding $\mathbf{\bar{e}}$, and processed with $1-D$ CNN layer to predict the duration of the de-duplicated tokens. Denoting the target and predicted durations by $\mathbf{d}=\{d_1, d_2,\dots, d_{T'}\}$ and $\mathbf{\hat{d}}=\{\hat{d_1}, \hat{d_2}, \hat{d_3}, \dots, \hat{d_{T'}}\}$ respectively, the loss for the duration predictor is defined as:
\begin{equation}\label{eq:lossdur}
    \mathrm{L}_{mse}^{dur} = \frac{1}{T'}||\mathbf{d} - \mathbf{\hat{d}} ||_{L_2} ^2
\end{equation}
\subsubsection{\textbf{Training Paradigm}}
\begin{figure}[t!]
    \centering
\includegraphics[width=0.47\textwidth,trim={1cm 5.5cm 2.5cm 3cm},clip]{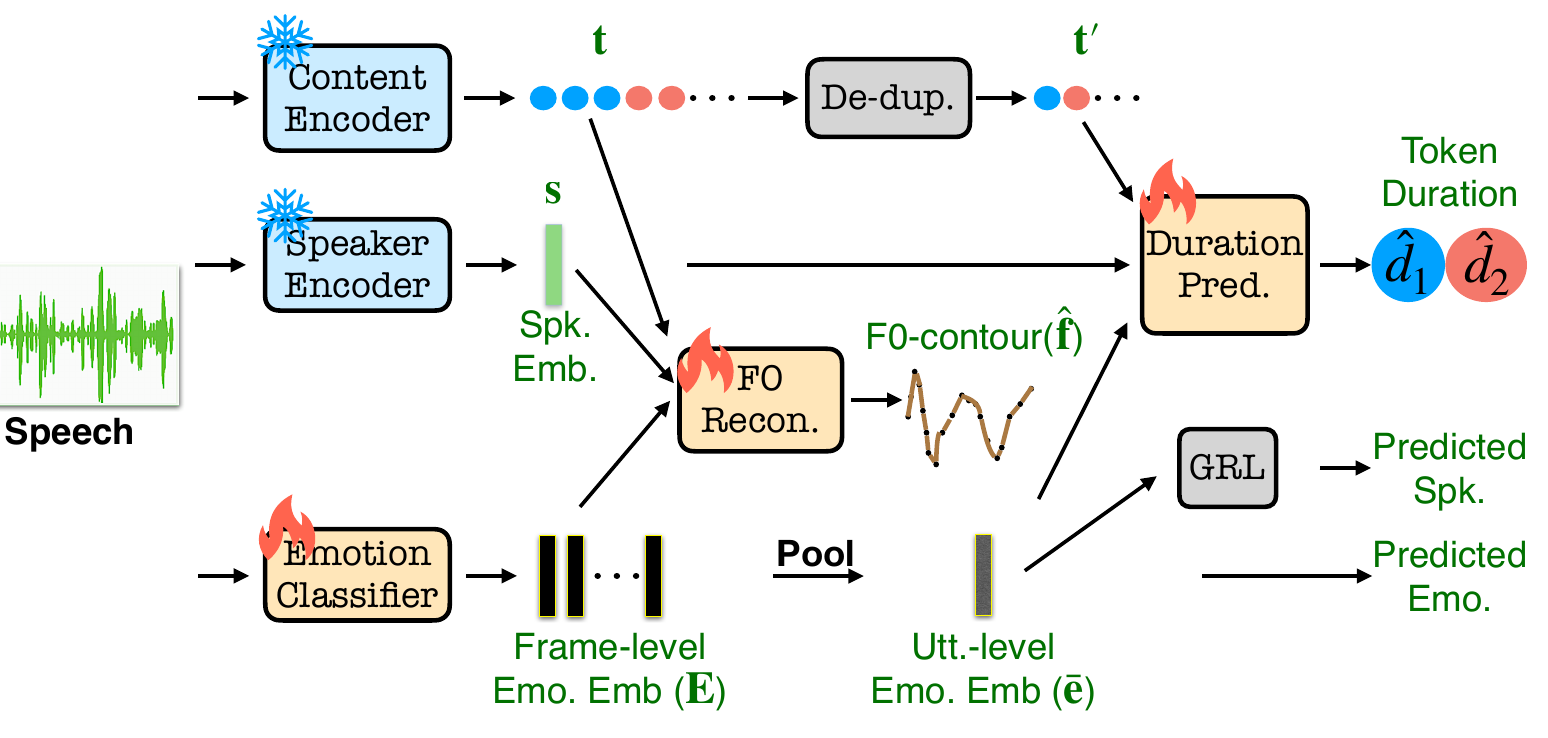}

    \caption{The different  factors that are derived from the speech in the analysis phase. The emotion classifier is trained with a speaker adversarial loss. The frame-level embeddings ($\mathbf{E}$), the speaker embedding ($\mathbf{s}$) and speech tokens ($\mathbf{t}$) are used to reconstruct the pitch contour ($\mathbf{\hat{f}}$). Further, the utterance-level emotion embedding ($\mathbf{\bar{e}}$) is used along with the de-duplicated tokens $\mathbf{t}^{'} = \{t_1,...,t_{T^{'}}\}$ to predict the duration of each of the tokens ($\mathbf{\hat{d}}$). All the blue blocks are kept frozen while the yellow blocks are trained. Grey blocks do not contain any learnable parameters.}
    \label{fig:emo_entire}

\end{figure}

The duration predictor, emotion classifier, and pitch contour reconstruction modules are trained jointly. The total loss  is:
\begin{equation}\label{eq:losstot}
    \mathrm{L}_{all} = \lambda_{e}\mathrm{L}_{tot-emo} + \lambda_{f}\mathrm{L}^{f0} + \lambda_{d}\mathrm{L}_{mse}^{dur}
\end{equation}
where $\lambda_{e}$, $\lambda_{f}$ and $\lambda_{d}$ are weighting coefficients.\\
The extraction of these components is shown in Fig.~\ref{fig:emo_entire}.
\subsection{Speech Reconstruction}
\label{sec:synth}
The token sequence, 
$\mathbf{t}$, 
is vectorized using a learnable embedding layer.  
Separately, the reconstructed pitch contour of the speech signal $\mathbf{\hat{f}}$ is vectorized through a learnable CNN-BiLSTM  network. These are concatenated with the vectorized token sequence, speaker embedding $\mathbf{s}$, and the utterance-level emotion embedding $\mathbf{\bar{e}}$ and are fed to the BigVGAN~\cite{leebigvgan} model for speech synthesis.

The BigVGAN network was proposed  as a universal vocoder that can generate \textcolor{black}{high-fidelity} raw speech waveform~\cite{leebigvgan}. The model was also shown to generalize well for various out-of-distribution scenarios without fine-tuning. 
The BigVGAN model upsamples the input sequence $N_w$ times using convolutional layers with residual connections.  
This network, called the generator, employs   \textit{Snake} activations ($f_\alpha(x) = x + \frac{1}{\alpha}\sin^2(\alpha x)$) for introducing the periodicity in speech signals~\cite{leebigvgan}. The discriminator network consists of two types - 1) Multi Period Discriminator (MPD)  to classify real from generated speech samples by focusing on  periodic structures of speech and 2) Multi Resolution Discriminator (MRD), where a number of discriminators are trained to separate real from generated speech samples by operating on spectrograms at different  resolutions. We use the same loss functions and hyper-parameters that are mentioned in Lee et al.~\cite{leebigvgan}. 
Thus, the reconstructed speech signal $\mathbf{\hat{x}} = \{\hat{x_1}, \hat{x_2}, \hat{x_3}, \dots, \hat{x_N}\}$ is derived from the BigVGAN as, $
    \mathbf{\hat{x}} = \mathtt{BigVGAN}(\mathbf{t}, \mathbf{s}, \mathbf{\bar{e}}, \mathbf{\hat{f}})$
\begin{figure}[t!]
    \centering
\includegraphics[width=0.47\textwidth,trim={0cm 5cm 0cm 2cm},clip]{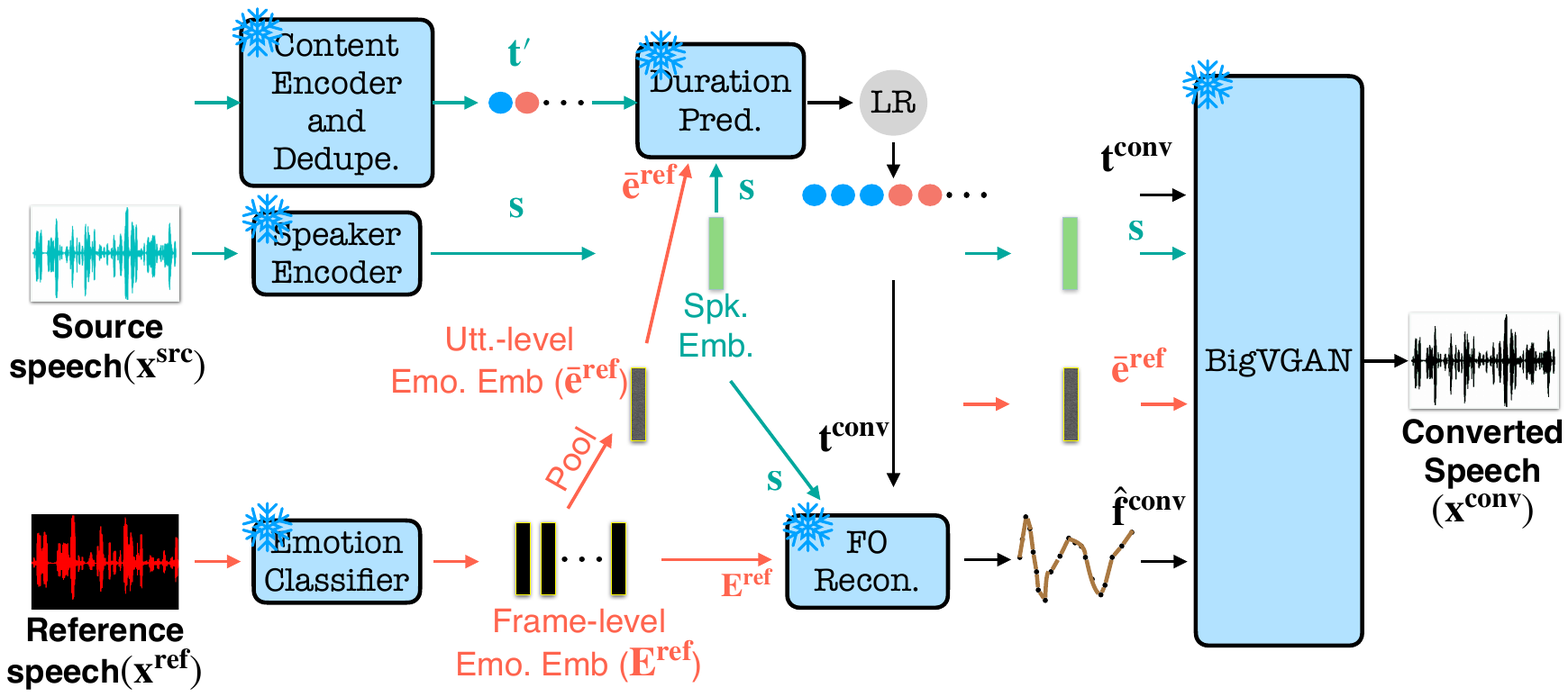}

    \caption{ Emotional style transfer - The frame-level ($\textbf{E}^{ref}$) and utterance-level ($\mathbf{\bar{e}}^{ref}$) embeddings are extracted from the reference speech. The duration prediction is performed using source tokens $\mathbf{t}^{'}$, speaker vector $\mathbf{s}$ and emotion embeddings
    $\mathbf{\bar{e}}^{ref}$. These predicted durations $\mathbf{\hat{d}}^{conv}$ are used to generate duplicated token sequence $\mathbf{t}^{conv}$. With this token sequence, $\mathbf{E}^{ref}$ and the speaker embedding $\mathbf{s}$, the $F_0$ contour is predicted, $\mathbf{\hat{f}}^{conv}$. Finally, the token sequence, speaker and emotion embeddings, and the predicted F0 contour are passed to the BigVGAN model to generate the converted speech.}
    \label{fig:conversion}

\end{figure}
\subsection{Emotion Style Transfer}
\label{sec:emoconv}
The emotion style transfer consists of the following steps:
\begin{itemize}
    \item \textbf{Emotion embeddings}: The frame-level and utterance-level emotion embeddings are extracted from the reference speech $\mathbf{x^{ref}}$, denoted as $\mathbf{E^{ref}}$ and $\mathbf{\bar{e}^{ref}}$.  
    \item \textbf{Token sequence}: We extract the de-duplicated token sequence $\mathbf{t}^{'}$ and the speaker embedding $\mathbf{s}$ from the source speech. Next, we predict the duration of each de-duplicated token using the emotion embedding $\mathbf{\bar{e}^{ref}}$ (similar to Eq.~\ref{eq:dur})  as:
    \begin{equation}\label{eq:dur_pred}
    \mathbf{\hat{d}}^{conv} = \mathtt{D_{pred}}(\mathbf{\mathbf{t}^{'}}, \mathbf{s}, \mathbf{\bar{e}^{ref}}) 
\end{equation} 
    The tokens in the sequence $\mathbf{t}^{'}$ are
    duplicated using $\mathbf{\hat{d}}^{conv}$,
    \begin{equation}
    \mathbf{t}^{conv} = \texttt{dup}(\mathbf{t}^{'},\mathbf{\hat{d}}^{conv}).     
    \end{equation} 
    
    \item \textbf{F0 Contour}: The token sequence $\mathbf{t}^{conv}$ is  vectorized to form $\mathbf{C}^{conv}$ and combined with     embeddings $\mathbf{s}$, $\mathbf{E^{ref}}$, to predict the pitch-contour (similar to Eq.~\ref{eq:f0}) as,
    \begin{equation}\label{eq:f0_pred}
    \mathbf{\hat{f}^{conv}} = \texttt{Attn}(\mathbf{C}^{conv},~~ \mathbf{s}+ \mathbf{E^{ref}},~~\mathbf{s}+ \mathbf{E^{ref}} ) 
\end{equation}

    \item \textbf{Speech Synthesis}: Finally, the converted speech is, 
    \begin{equation}
    \mathbf{x^{conv}} = \mathtt{BigVGAN}(\mathbf{t^{conv}}, \mathbf{s}, \mathbf{\bar{e}^{ref}}, \mathbf{\hat{f}^{conv}}) 
\end{equation}
    \end{itemize}
These steps are highlighted in Fig.~\ref{fig:conversion}.

\section{Experiments and Results}
\subsection{Datasets and Pretraining}
\label{dataset}
For the content encoder, we use the soft-HuBERT base model as proposed by Niekerk et al.~\cite{van2022comparison}, which was pre-trained on $960$ hours of the LibriSpeech dataset~\cite{panayotov2015librispeech}.  
We train a k-means clustering model on randomly selected $10\%$ of the training dataset, which forms the speech tokenizer.

The emotion embeddings are extracted from an emotion classifier model trained on the Emotional Speech Database (ESD)~\cite{zhou2021seen}. The ESD database consists of $350$ parallel utterances spoken by $10$ native English and $10$ native Chinese speakers and covers $5$ emotion categories (neutral, happy, angry, sad, and surprise). We only use the English training subset of the ESD dataset to build the emotion classifier model. We follow the dataset's predefined train-validation-test splits, using $300$ utterances per speaker per emotion for training. This results in a total of $15000$ training utterances, with $2500$ unseen utterances ($50$ per speaker per emotion) for validation.

The speaker encoder is initialized with the ECAPA-TDNN model~\cite{desplanques20_interspeech}, which was pre-trained on $2794$ hours and $7363$ speakers from the VoxCeleb dataset. 
We fine-tune the speaker encoder model using the emotion-adversarial loss (Eq.~\ref{eq:lossspkr}) on the ESD dataset. 

\textcolor{black}{Unless otherwise stated, all the modules are trained on the gender-balanced English training partition of ESD. To analyze the impact of training data diversity on generalization, we additionally train a variant (denoted as \method{}-diverse) using the original ESD data augmented with $100$ hours of LibriSpeech. For this variant, the speech tokenizer, speaker encoder, and BigVGAN model are trained on the augmented dataset. This variant is used solely to study scaling behavior under increased speaker variability.}

\begin{table*}[t]
\centering
\caption{Various evaluation settings for emotion style transfer.}
\label{tab:evaluation-settings}
\resizebox{0.95\textwidth}{!}{
\begin{tabular}{@{} l|l|l|l|cccc|c @{}}
\toprule
\textbf{Setting} & \textbf{Description} & \textbf{Src. Dataset} & \textbf{Ref. Dataset} & \textbf{Matched Spkr.} & \textbf{Matched Text} & \textbf{Seen Emo.} & \textbf{Seen Spkr.} & \textbf{\#Samples} \\
\midrule
SSST & Same Speaker Same Text & ESD & ESD & \Checkmark & \Checkmark & \Checkmark & \Checkmark & $1200$ \\
SSDT & Same Speaker Different Text & ESD & ESD & \Checkmark & \XSolidBrush   & \Checkmark & \Checkmark & $1160$ \\
DSST & Different Speaker Same Text & ESD & ESD & \XSolidBrush   & \Checkmark & \Checkmark & \Checkmark & $10800$ \\
DSDT & Different Speaker Different Text & ESD & ESD & \XSolidBrush   & \XSolidBrush   & \Checkmark & \Checkmark & $10440$ \\ \midrule
UTE  & Unseen Target Emotions & ESD & CREMA-D & \XSolidBrush & \XSolidBrush & \XSolidBrush   & \Checkmark & $1000$ \\
USS  & Unseen Source Speakers & TIMIT & ESD & \XSolidBrush   & \XSolidBrush   & \Checkmark & \XSolidBrush & $800$ \\
USUE  & Unseen Speaker Unseen Emotion & TIMIT & CREMA-D & \XSolidBrush   & \XSolidBrush   & \XSolidBrush & \XSolidBrush & $2000$ \\
\bottomrule
\end{tabular}}
\end{table*}
\subsection{Implementation}
The content encoder utilizes soft-HuBERT representations which are quantized using a k-means clustering algorithm with $K=100$ clusters~\cite{polyak2021speech}. The speaker encoder is trained for $10$ epochs with $\lambda_{adv}^{emo} = 10$ (Eq.~\ref{eq:lossspkr}) and a batch size of $32$. The emotion classifier is trained with $\lambda_{adv}^{spk} = 1$ (Eq.~\ref{eq:lossemo}) and is jointly optimized with the pitch reconstruction network and duration predictor using the total loss function in Eq.~\ref{eq:losstot}. The weighting coefficients are set as $\lambda_{e} = 1000$, $\lambda_{f} = 1$, and $\lambda_{d} = 10$, where the values were set based on  validation loss. 
This joint training process is performed for $200$ epochs with a batch size of $32$ and a learning rate of $1e-4$. During inference, the predicted durations for the tokens are constrained to lie within $\pm40\%$ of the original source token durations. 
 \textcolor{black}{This constraint is applied as a post-processing step to prevent extreme duration deviations that may result in unnatural rhythm or unstable pitch reconstruction. We observed that,  without this constraint, occasional outlier predictions can lead to excessive stretching or compression of segments. The $\pm40\%$ bound was selected empirically to balance expressive variation and speech naturalness.}
The cross-attention model in the pitch reconstruction module has $4$ attention heads and a hidden dimension of $256$, while the duration predictor consists of a $1$D-CNN network with a kernel size of $3$ and a hidden dimension of $256$.
Finally, the speech reconstruction module, BigVGAN~\cite{leebigvgan}, is trained with a batch size of $16$ and a learning rate of $1e-4$. 
All models are optimized using the AdamW optimizer~\cite{loshchilov2017decoupled}. \textcolor{black}{For the \method{}-diverse variant, the same architectural configuration and hyperparameters are used, with the training performed on the augmented dataset described in Section~\ref{dataset}.} The code and samples are open-sourced.\footnote{\url{https://github.com/iiscleap/A2A-ZEST}}.

\subsection{Evaluation Settings}
\label{sec:evaluation}
There are two broad evaluation settings,
\begin{itemize} 
    \item \textbf{Seen}: The source speech contains unseen content from the test data of ESD, but the speakers and emotions are present in the training data.
    \item \textbf{Unseen}: Either the source speaker or the reference speech emotion or both are unseen during training.
\end{itemize}
The different test evaluation settings are outlined in Table~\ref{tab:evaluation-settings}. For the four evaluation settings where both source and reference speech are drawn from ESD~\cite{zhou2022emotional}, the source speech is always neutral, and the reference speech belongs to one of the four emotion categories: ``happy'', ``angry'', ``sad'', and ``surprised''. In the SSDT and DSDT settings, we randomly select $10$ neutral utterances (one per speaker) from ESD to serve as source speech.\\
For unseen evaluation settings, we use CREMA-D for unseen emotions and TIMIT for unseen source speakers. In the Unseen Target Emotion (UTE) setting, we select 50 utterances each from the “fear” and “disgust” categories in CREMA-D as reference speech, paired with 10 neutral utterances (one per speaker) from ESD, yielding 1000 evaluation utterances. In the Unseen Source Speaker (USS) setting, 100 utterances from TIMIT are used as source speech, each paired with 8 emotional utterances from ESD (2 per emotion, excluding neutral), resulting in 800 evaluation utterances. 
In the Unseen Speaker Unseen Emotion (USUE) setting, we pair the same 100 TIMIT source utterances with 10 utterances each from the “fear” and “disgust” categories in CREMA-D as reference speech, for a total of 2000 evaluation utterances.

\begin{table*}[t!]
\caption{Objective evaluation results. Here, WER - Word Error Rate, Emo. Sim. - Emotion Similarity according to emotion2vec embeddings, Spk.-Sim. - Speaker Similarity according to the Resemblyzer embedding. Word PCC refers to the word speaking rate Pearson correlation coefficient. Different test settings are described in Table~\ref{tab:evaluation-settings}. $*$ indicates that these entries are not \textbf{marked in bold} or \underline{underline} as StarGANv2-EST is not applicable to $2$ out of the $7$ test settings.}
\label{tab:obj results}
\centering
\renewcommand{\arraystretch}{1.2}
\resizebox{\textwidth}{!}{
\begin{tabular}{@{}l|l|l|ccccccc|c@{}}
\toprule
\textbf{Category} & \textbf{Method} & \textbf{Metric} & SSST & SSDT & DSST & DSDT & UTE & USS & USUE & Avg \\
\midrule
\multirow{5}{*}{\textbf{Emotion Transfer}}
& StarGANv2-EST~\cite{li2021starganv2} &  & $0.39$ & $0.37$ & $0.37$ & $0.37$ & - & $0.25$ & - & $0.35$ \\
\cmidrule{2-2}\cmidrule{4-11}
& VEVO~\cite{zhang2025vevo}  &\multirow{5}{*}{Emo.-Sim. ($\uparrow$)}  & $-0.01$ & $-0.03$ & $-0.01$ & $-0.04$ & $-0.01$ & $0.37$ & $-0.02$ & $0.04$ \\
\cmidrule{2-2}\cmidrule{4-11}
& ZEST~\cite{dutta2024zero} &  & $0.59$ & $0.62$ & $0.52$ & $0.51$ & $0.53$ & $0.48$ & $0.41$ & $0.52$ \\
\cmidrule{2-2}\cmidrule{4-11}
& \method{} &  & $\mathbf{0.69}$ & $\mathbf{0.71}$ & $\mathbf{0.58}$ & $\mathbf{0.56}$ & $\mathbf{0.64}$ & $\mathbf{0.59}$ & $\mathbf{0.51}$ & $\mathbf{0.61}$ \\
\cmidrule{2-2}\cmidrule{4-11}
& \method{}\textcolor{black}{-diverse} &  & \textcolor{black}{$\underline{0.67}$} & \textcolor{black}{$\underline{0.70}$} & \textcolor{black}{$\underline{0.55}$} & \textcolor{black}{$\underline{0.52}$} & \textcolor{black}{$\underline{0.59}$} & \textcolor{black}{$\underline{0.57}$} & \textcolor{black}{$\underline{0.50}$} & \textcolor{black}{$\underline{0.59}$} \\
\midrule

\multirow{5}{*}{\textbf{Rhythm Transfer}}
& StarGANv2-EST~\cite{li2021starganv2} &  & $0.54$ & $0.03$ & $0.43$ & $0.02$ & - & $0.02$ & - & $0.21$ \\
\cmidrule{2-2}\cmidrule{4-11}
& VEVO~\cite{zhang2025vevo} & \multirow{5}{*}{Word PCC ($\uparrow$)} & $\underline{0.65}$ & $0.10$ & $\mathbf{0.59}$ & $\mathbf{0.16}$ & $0.01$ & $0.08$ & $0.01$ & $0.23$ \\
\cmidrule{2-2}\cmidrule{4-11}
& ZEST~\cite{dutta2024zero} & & $0.53$ & $0.04$ & $0.44$ & $4\times10^{-4}$ & $5\times10^{-3}$ & $0.02$ & $2\times10^{-4}$ & $0.15$ \\
\cmidrule{2-2}\cmidrule{4-11}
& \method{} &  & $\mathbf{0.68}$ & $\mathbf{0.18}$ & $\underline{0.58}$ & $\underline{0.14}$ & $\mathbf{0.07}$ & $\mathbf{0.11}$ & $\mathbf{0.06}$ & $\mathbf{0.26}$ \\
\cmidrule{2-2}\cmidrule{4-11}
& \method{}\textcolor{black}{-diverse} &  & \textcolor{black}{$\mathbf{0.68}$} & \textcolor{black}{$\underline{0.16}$} & \textcolor{black}{$\underline{0.58}$} & \textcolor{black}{$0.12$} & \textcolor{black}{$\underline{0.05}$} & \textcolor{black}{$\underline{0.09}$} & \textcolor{black}{$\underline{0.04}$} & \textcolor{black}{$\underline{0.25}$} \\
\midrule \midrule

\multirow{5}{*}{\textbf{Content Preservation}} 
& StarGANv2-EST~\cite{li2021starganv2} &  & $5.72$ &$8.28$ & $6.61$ & $\underline{6.02}$ & - & $\mathbf{9.41}$ & - & $7.21^{*}$ \\
\cmidrule{2-2}\cmidrule{4-11}
& VEVO~\cite{zhang2025vevo} & \multirow{5}{*}{WER ($\downarrow$)} & $\underline{4.55}$ & $7.65$ & $\underline{4.72}$ & $7.68$ & $6.16$ & $14.36$ & $14.06$ & $\underline{8.45}$ \\
\cmidrule{2-2}\cmidrule{4-11}
& ZEST~\cite{dutta2024zero} &  & $5.11$ & $5.32$ & $5.19$ & $6.11$ & $\underline{5.92}$ & $16.57$ & $16.42$ & $8.66$ \\
\cmidrule{2-2}\cmidrule{4-11}
& \method{} &  & $5.58$ & $\underline{4.94}$ & $5.55$ & $6.48$ & $6.03$ & $13.53$ & $\underline{12.43}$ & $\underline{7.79}$ \\
\cmidrule{2-2}\cmidrule{4-11}
& \method{}\textcolor{black}{-diverse} &  & $\textcolor{black}{\mathbf{3.65}}$ & \textcolor{black}{$\mathbf{3.15}$} & \textcolor{black}{$\mathbf{3.51}$} & \textcolor{black}{$\mathbf{4.82}$} & \textcolor{black}{$\mathbf{5.02}$} & \textcolor{black}{$\underline{11.63}$} & \textcolor{black}{$\mathbf{10.84}$} & \textcolor{black}{$\mathbf{6.09}$} \\
\midrule

\multirow{5}{*}{\textbf{Speaker Preservation}}
&  StarGANv2-EST~\cite{li2021starganv2}&  & $\underline{0.76}$ & $\underline{0.76}$ & $0.67$ & $0.65$ & - & $\underline{0.65}$ & - & $0.70$ \\
\cmidrule{2-2}\cmidrule{4-11}
&  VEVO~\cite{zhang2025vevo}& \multirow{5}{*}{Spk.-Sim.($\uparrow$)} & $\mathbf{0.86}$ & $\mathbf{0.87}$ & $\mathbf{0.86}$ & $\mathbf{0.86}$ & $\mathbf{0.85}$ & $\mathbf{0.87}$ & $\mathbf{0.86}$ & $\mathbf{0.86}$ \\
\cmidrule{2-2}\cmidrule{4-11}
& ZEST~\cite{dutta2024zero} & & $0.74$ & $0.75$ & $0.74$ & $0.72$ & $0.73$ & $0.54$ & $0.53$ & $0.68$ \\
\cmidrule{2-2}\cmidrule{4-11}
&  \method{} &  & $0.74$ & $0.73$ & $0.74$ & $0.72$ & $0.73$ & $0.53$ & $0.53$ & $0.68$ \\
\cmidrule{2-2}\cmidrule{4-11}
& \method{}\textcolor{black}{-diverse} &  & \textcolor{black}{$0.75$} & \textcolor{black}{$0.74$} & \textcolor{black}{$\underline{0.76}$} & \textcolor{black}{$\underline{0.74}$} & \textcolor{black}{$\underline{0.76}$} & \textcolor{black}{$0.61$} & \textcolor{black}{$0.59$} & \textcolor{black}{$\underline{0.71}$} \\
\bottomrule
\end{tabular}}
\end{table*}

\subsection{Objective Evaluation Metrics} \label{sec:metrics}

\begin{itemize}
    \item \textbf{Emotion transfer w.r.t reference}: We extract embeddings from emotion2vec~\cite{ma2024emotion2vec} for both the reference and converted speech and compute their mean cosine similarity (denoted by Emo.-Sim.).
    \item \textbf{Rhythm transfer w.r.t reference}: We force align the reference and the converted speech signals with their corresponding text transcripts using a pre-trained ASR model~\cite{baevski2020wav2vec}. Following this, the word speaking rate is computed for the reference and the converted speech. The Pearson Correlation Coefficient (PCC) between the reference speaking rate and the converted speaking rate is used as measure of rhythm transfer. This is similar to the metric proposed by Barrault et al.~\cite{barrault2023seamless}.
    \item \textbf{Content preservation w.r.t source}: We transcribe the converted speech using an automatic speech recognition (ASR) system. Specifically, we employ the pre-trained Whisper-large-v3 model~\cite{radford2023robust} for performing speech recognition.  The word error rate (WER) is measured using the ground-truth transcripts of the source speech.
    
    \item \textbf{Speaker preservation w.r.t source}: We report the average cosine similarity of speaker embeddings 
     between the source and the converted speech.
     To derive these speaker embeddings, we use the Resemblyzer tool~\footnote{\url{https://github.com/resemble-ai/Resemblyzer}}. This is denoted as Spk.-Sim.    
\end{itemize}

\subsection{Comparison with baselines}
Three baseline systems are considered:
\begin{itemize}
    \item \textbf{StarGANv2-EST}: Li et al.~\cite{li2021starganv2} proposed a StarGAN-v2 network for voice conversion. An auxiliary classifier was trained to classify the source speaker alongside the discriminator, enabling speaker conversion. For our experiments, we modify this architecture to perform emotion transfer by using the auxiliary network to classify emotions instead. We refer to this modified model as StarGANv2-EST. It is trained on the same dataset as \method{}. Note that this baseline cannot handle emotions unseen during training, and is therefore not evaluated on the UTE and USUE test settings.
    \item \textbf{VEVO}: We use the pre-trained VEVO model proposed by Zhang et al.~\cite{zhang2025vevo}. Since this model was trained on a much larger corpus than \method{}, we do not fine-tune it for our test settings. Instead, we use VEVO’s style imitation pipeline for comparison.
    \item \textbf{ZEST}: We also compare with  prior work - ZEST~\cite{dutta2024zero}.
\end{itemize}

\subsection{Objective Results}\label{sec:obj_results}

The results for these objective tests are shown in Table \ref{tab:obj results}. 
The following are the insights drawn from these  evaluations.
\begin{itemize}

    \item \textbf{Emotion Transfer}: \method{} achieves the highest emotion similarity across both seen and unseen settings. For seen speakers and emotions, it achieves $0.69$ similarity in the SSST setting and $0.71$ in SSDT, outperforming all baselines. In the unseen target emotion (UTE) scenario, the \method{} achieves $0.64$ similarity, demonstrating effective transfer of emotions, not observed during training. For unseen source speakers (USS), the model achieves $0.59$, demonstrating robust generalization to novel speakers. In the most challenging unseen speaker and unseen emotion (USUE) setting, \method{} attains $0.51$, where both the speaker and emotion are unseen. In contrast, StarGANv2-EST and other prior-works achieve low similarity scores, indicating their limited ability for effective emotion style transfer. \textcolor{black}{The \method{}-diverse exhibits similar emotion similarity on average ($0.59$ vs.\ $0.61$).}

    \item \textbf{Rhythm transfer}: For same-text settings, \method{} achieves $0.68$ correlation in SSST and $0.58$ in DSST, showing strong rhythm transfer. In different-text settings, SSDT and DSDT, the correlation drops to $0.18$ and $0.14$, respectively, reflecting lower alignment due to text differences. A reduced correlation of $0.11$, $0.07$ and $0.06$ are observed for USS, UTE and USUE settings. However, across all settings, on average, \method{} ($0.26$) outperforms ZEST ($0.15$) and StarGANv2-EST ($0.21$), while its performance is similar to VEVO ($0.23$). \textcolor{black}{The \method{}-diverse variant maintains similar rhythm transfer performance.}

    \item \textbf{Content Preservation}: \method{} demonstrates consistent WER across test settings where the source speech is from ESD (i.e., all settings except USS and USUE). Although VEVO is trained on $60$K hours of speech, \method{} achieves lower WER in five out of the seven applicable test settings (SSDT, DSDT, UTE, USS, USUE). While StarGANv2-EST attains the lowest WER in some individual settings (DSDT and USS), it is not applicable for UTE and USUE, limiting its overall evaluation.
Overall, \method{} achieves the best average WER ($7.79$), demonstrating strong content preservation while performing emotion style transfer. \textcolor{black}{The \method{}-diverse variant substantially improves content preservation across all test settings, reducing the average WER from $7.79$ to $6.09$. Notably, improvements are also  observed in unseen-speaker conditions (USS: $13.53 \rightarrow 11.63$, USUE: $12.43 \rightarrow 10.84$), indicating enhanced linguistic stability.}

    \item \textbf{Speaker identity preservation}:Across all test settings, VEVO consistently achieves the highest speaker similarity, ranging from $0.85$ to $0.87$, demonstrating its strong ability to preserve speaker identity. \textcolor{black}{  When trained with additional speaker diversity, \method{}-diverse improves speaker similarity consistently across all settings, particularly in unseen-speaker scenarios (USS: $0.53 \rightarrow 0.61$, USUE: $0.53 \rightarrow 0.59$). These results suggest that speaker preservation in unseen conditions is influenced by training data diversity. Training with more speakers and with diverse emotional speech will form part of our future scope of work.}
    
\end{itemize}

\begin{figure}[t!]
    \centering
    \includegraphics[width=0.5\textwidth,trim={1cm 6cm 6cm 3cm},clip]{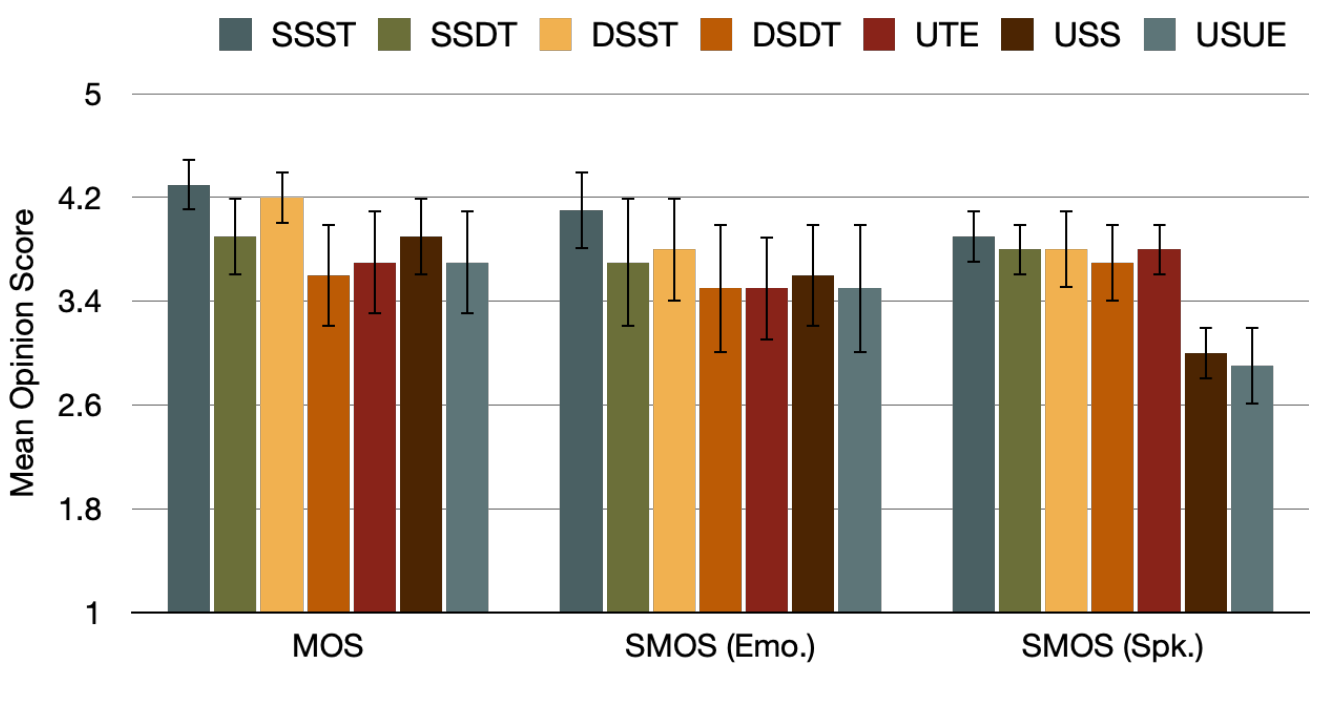}
    \caption{Subjective evaluation on the different test settings. Abbreviations used: MOS- Mean Opinion Score, SMOS - Similarity Mean Opinion Score. \textcolor{black}{The $95\%$ confidence intervals are also shown.}}
     \label{fig:subjective}
\end{figure}
\begin{figure}[t!]
    \centering
    \includegraphics[width=0.5\textwidth,trim={1cm 4cm 9cm 4cm},clip]{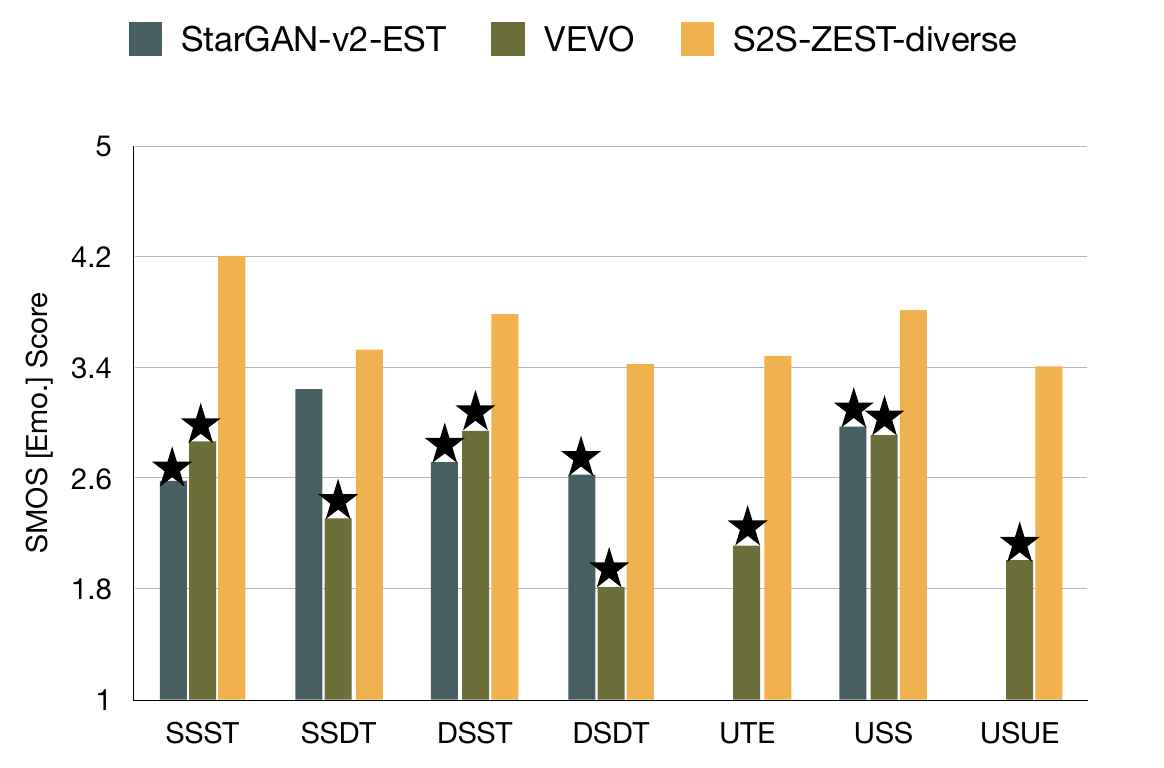}
    \caption{Subjective evaluation on the different test settings. SMOS stands for Similarity Mean Opinion Score. {\small \FiveStar}~indicates that the difference in scores between the baseline and \method{}\textcolor{black}{-diverse} is statistically significant (p $<$ $0.05$)}
     \label{fig:subjective_comp}
\end{figure}
\subsection{Subjective tests}
\subsubsection{Evaluation of \method{}\textcolor{black}{-diverse}}
We conduct listening tests using the Prolific\footnote{\url{https://www.prolific.co}} platform.
We recruited $30$ participants to perform the subjective evaluation. 
We chose $44$  recordings, with $8$ recordings from each of the $4$ test settings (SSST, SSDT, DSST, DSDT) and $4$ recordings each from UTE, USS, and USUE settings. The recordings were presented in a random order. The participants were also provided with training examples to clarify the objective of the test.
 
All the participants in the survey were asked to give their opinion score on the speech files (range of $1$-$5$) based on three criteria - i) Emotion similarity between the converted and the reference signal, ii) Quality of the converted speech, and iii) Speaker similarity between the converted and the source signal. 
The subjective evaluation results (in terms of mean opinion score (MOS)) are reported in Figure~\ref{fig:subjective}. The key observations from the subjective tests are as follows:
\begin{itemize}
    \item  The performance of \method{}\textcolor{black}{-diverse} is best on all the three criteria for the SSST test setting. This is expected as transfer of emotional speaking style from the reference to the source, when they share the same content and speaker, is the easiest of all the test settings. 
    \item When the source and reference share the same content but have different speakers (DSST), subjective scores indicate that emotion transfer is easier compared to the SSDT setting (same speaker, different content).
    \item The SMOS (Emo.) scores in the UTE and USUE setting are comparable to the DSDT setting, indicating that generalizing to unseen emotions is not any more challenging compared to transferring seen emotions across different speakers and content. 
    \item  \textcolor{black}{The USS and USUE settings exhibit lower speaker similarity scores compared to seen-speaker conditions, reflecting the inherent difficulty of preserving speaker identity for speakers unseen during training. While the diverse variant is a step in this direction, training the encoders and the vocoder with more speaker data will help improve the speaker generalization further.} 

\end{itemize}
\subsubsection{Comparison of \method{}\textcolor{black}{-diverse} with baselines}
We consider $4$ converted files from each of the $7$ test settings, and for each of the three methods (StarGANv2-EST does not have any outputs for the UTE and USUE test settings), and ask \textcolor{black}{$30$ participants} to rate the emotion speaking style similarity between the reference speech and the converted speech on a scale of $1$ to $5$. The results for this subjective test are shown in Figure~\ref{fig:subjective_comp}. 

The results indicate that \method{}\textcolor{black}{-diverse} outperforms both the baselines by a significant margin for all the test settings (except for StarGANv2-EST for the SSDT test setting). This showcases the superiority of \method{}\textcolor{black}{-diverse} in emotional style transfer compared to the baseline methods and follows the trends observed in the objective evaluations. 
\begin{figure}[t!]
    \centering
    \includegraphics[width=0.5\textwidth,trim={1cm 7cm 11cm 5.5cm},clip]{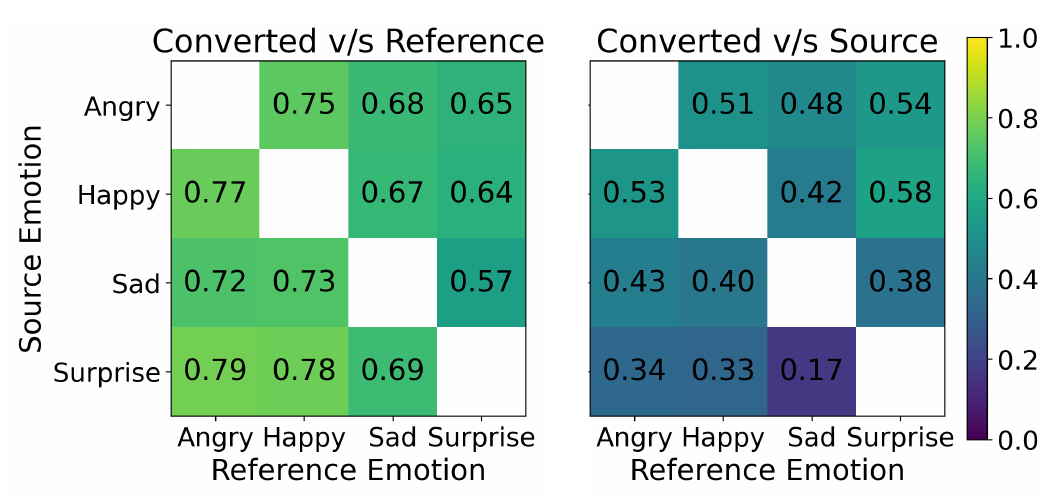}
    \caption{\textcolor{black}{Average emotion similarity between different pairs of emotions.}}
     \label{fig:any2any}
\end{figure}
\subsection{\textcolor{black}{Any-to-Any Emotion Style Transfer}}
\textcolor{black}{ In this experiment, the source and reference utterances share identical speaker identity and linguistic content, differing only in emotional category. This controlled setup is constructed from the ESD dataset and results in $3600$ test utterances.}

\textcolor{black}{We evaluate all $4 \times 4$ source–target emotion combinations among \{angry, happy, sad, surprise\}. Emotion similarity is computed using the metric  described in Sec.~\ref{sec:metrics}. Fig.~\ref{fig:any2any} presents two emotion similarity matrices: (i) similarity between the converted speech and the reference emotion, and (ii) similarity between the converted speech and the original source emotion. Successful emotion transfer should increase similarity w.r.t. the reference while reducing similarity w.r.t. the source.} 

\textcolor{black}{Across all source–target pairs, the converted speech consistently exhibits high similarity w.r.t. the reference emotion (e.g., Surprise$\rightarrow$Angry: $0.79$, Angry$\rightarrow$Happy: $0.75$). At the same time, similarity w.r.t. the original source emotion is comparatively lower for most pairs, suggesting that the model has moved away from the source emotion.}
\textcolor{black}{For certain transitions (e.g., Sad$\rightarrow$Surprise), the reference similarity is relatively lower, reflecting the increased difficulty of converting between these two emotional states. Overall, these results demonstrate that \method{}\textcolor{black}{-diverse} supports directional emotion style transfer across expressive source conditions, beyond neutral-source scenarios.}

\begin{table}[t]
\centering
\caption{\textcolor{black}{Robustness under additive MUSAN noise in the neutral-to-emotion setting.}}
\begin{tabular}{l|c|c}
\toprule
\textcolor{black}{Condition} & \textcolor{black}{Emo. Sim. $\uparrow$} & \textcolor{black}{WER (\%) $\downarrow$} \\
\midrule
\textcolor{black}{Clean} & \textcolor{black}{$0.67$} & \textcolor{black}{$3.65$} \\
\textcolor{black}{$20$ dB} & \textcolor{black}{$0.65$} & \textcolor{black}{$4.38$} \\
\textcolor{black}{$10$ dB} & \textcolor{black}{$0.64$} & \textcolor{black}{$7.74$} \\
\bottomrule
\end{tabular}
\label{tab:noise}
\end{table}
\subsection{\textcolor{black}{Robustness to Additive Noise}}
\textcolor{black}{
To evaluate robustness under noisy conditions, we conduct experiments in the SSST test setting (Table.~\ref{tab:evaluation-settings}) with additive MUSAN noise applied to the source speech at $20$ dB and $10$ dB SNR. The reference speech remains clean. Emotion similarity and WER are computed as described in Sec.~\ref{sec:metrics}. The results are summarized in Table~\ref{tab:noise}.}

\textcolor{black}{Emotion similarity remains largely stable under moderate noise, decreasing marginally from $0.67$ (clean) to $0.65$ ($20$ dB) and $0.64$ ($10$ dB).}
\textcolor{black}{However, in terms of intelligibility, WER increases progressively from $3.65\%$ (clean) to $4.38\%$ ($20$ dB) and $7.74\%$ ($10$ dB), highlighting the   impact of additive noise on content preservation. While recognition performance degrades at lower SNR levels, emotion similarity remains relatively stable, indicating that emotion style is effectively transferred despite corruption of the source signal by noise. }

\section{Discussion}
\textcolor{black}{In this section, all experiments are conducted using the base \method{} configuration trained on the ESD dataset.
}
\subsection{Choice of the speech tokenizer and synthesizer}
\begin{figure}[t!]
    \centering
\includegraphics[width=0.47\textwidth,trim={1.5cm 4cm 9.5cm 6cm},clip]{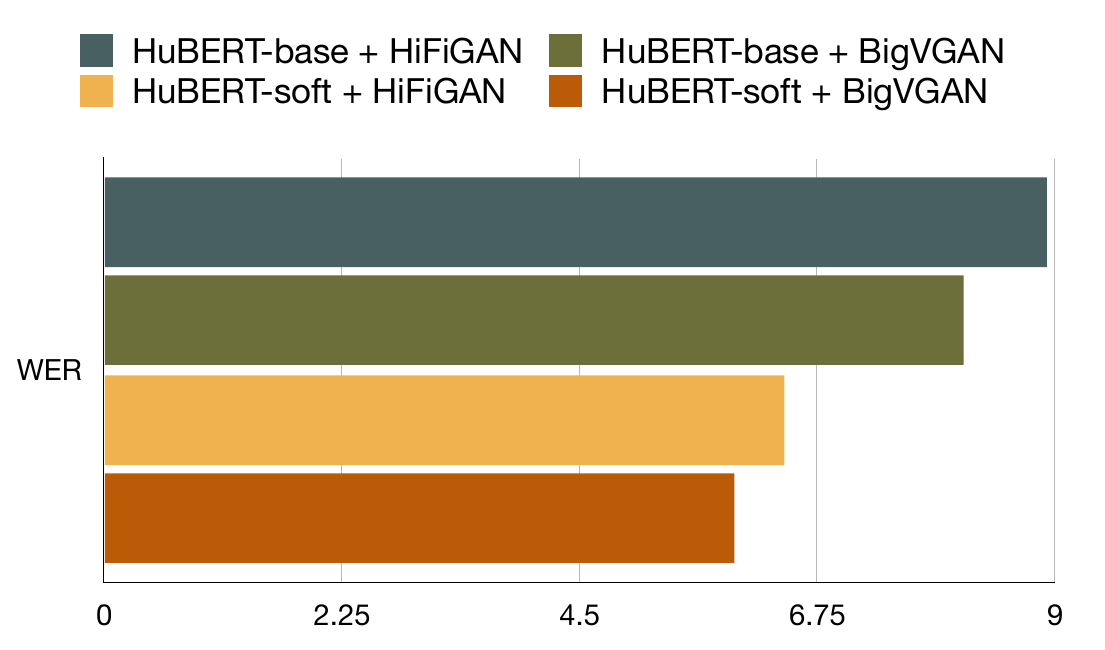}

     \caption{The WER of the \textcolor{black}{resynthesized} speech with various combinations of the speech tokenizer and synthesizer.}
    \label{fig:token_synth}

\end{figure}

To assess the impact of the speech tokenizer and synthesizer, we conduct an experiment where we replace the soft-HuBERT features with those from the pre-trained HuBERT-base model. We reconstruct the speech signals in the test set of ESD, with the tokens derived from HuBERT-base features. 
Additionally, we train a HiFi-GAN model as an alternative speech synthesizer with the tokens from the soft-HuBERT features. The WER results for the reconstructed speech signals with these different configurations are presented in Figure~\ref{fig:token_synth}.\\
\textbf{Key takeaways}: 1) Speech reconstructed using soft-HuBERT tokens achieves lower WER compared to HuBERT-base tokens, regardless of the synthesizer used. Prior work~\cite{van2022comparison} has shown that continuous soft-HuBERT features improve intelligibility over HuBERT-base tokens. Our results indicate that even after discretizing these features into tokens, speech reconstruction still achieves lower WER compared to HuBERT-base tokens, preserving this advantage. 2) The BigVGAN-base model consistently outperforms HiFi-GAN, irrespective of the tokenizer used, with the same number of parameters (14M). This improvement can be attributed to enhanced activation functions in BigVGAN.

\subsection{Impact of the duration predictor}\label{sec:dur_imp}
To evaluate the impact of the duration predictor, we conduct an experiment where we synthesize the converted speech using the source speech token sequence, without any duration modification.
These results are reported in Table~\ref{tab:dur_pred}.\\
\textbf{Key takeaways}: 1) Using the duration predictor leads to higher WER. However, it improves word-rate correlation, suggesting that better rhythm expression comes at the cost of reduced recognition on ASR trained with neutral speech. This aligns with observations by Maimon et al.~\cite{maimon2022speaking}. 2) The improvement of the emotion similarity across all the $4$ test settings indicates that inclusion of the duration predictor leads to effective emotion style transfer.
\begin{table}[t!]
\caption{WER, Emo.-Sim., and Pearson correlation coefficient (PCC) with respect to words for four test settings on the ESD dataset, using \method{} and \method{} without Dur. pred.}
\label{tab:dur_pred}
\centering
\renewcommand{\arraystretch}{1.2}
\setlength{\tabcolsep}{3pt} 
\resizebox{\linewidth}{!}{ 
\small
\begin{tabular}{@{}c|cc|cc|cc@{}}
\toprule
\multirow{2}{*}{\shortstack{Test \\ Setting}} 
& \multicolumn{2}{c|}{WER $\downarrow$} 
& \multicolumn{2}{c|}{Emo.-Sim. (\%) $\uparrow$} 
& \multicolumn{2}{c}{Word PCC $\uparrow$} \\  
\cmidrule(l){2-7} 
& \method{} & \shortstack{$-$Dur. \\ pred.}  
& \method{} & \shortstack{$-$Dur. \\ pred.}  
& \method{} & \shortstack{$-$Dur. \\ pred.}  \\ 
\midrule
SSST & $5.58$ & $\mathbf{4.76}$ & $\mathbf{0.69}$ & $0.61$ & $\mathbf{0.68}$ & $0.51$ \\
SSDT & $4.94$ & $\mathbf{4.10}$ & $\mathbf{0.71}$ & $0.62$ & $\mathbf{0.18}$ & $0.04$ \\
DSST & $5.55$ & $\mathbf{4.85}$ & $\mathbf{0.58}$ & $0.51$ & $\mathbf{0.58}$ & $0.41$ \\
DSDT & $6.48$ & $\mathbf{5.59}$ & $\mathbf{0.56}$ & $0.52$ & $\mathbf{0.14}$ & $5\times10^{-4}$ \\ 
\bottomrule
\end{tabular}
}
\end{table}

\begin{figure}[t!]
    \centering
\includegraphics[width=0.5\textwidth,trim={4cm 8.5cm 3cm 5cm},clip]{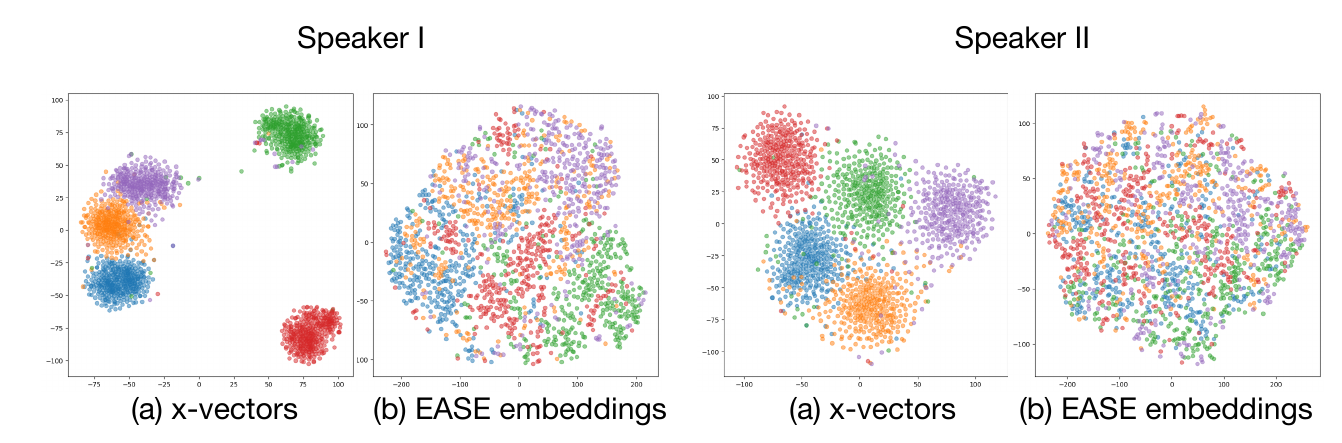}

    \caption{t-SNE plots for speaker embeddings with and without adversarial loss for two speakers I and II. In these figures, the colors correspond to $5$  emotion categories in ESD.}
    \label{fig:ease_ablation}

\end{figure}
\begin{table}[t!]
\caption{WER, Emo.-Sim., and Spk.-Sim. for four test settings on the ESD dataset, using \method{}, \method{} without Emo. Adv. loss ($\lambda_{adv}^{emo}=0$), and \method{} without Spk. Adv. loss ($\lambda_{adv}^{spk}=0$)}
\label{tab:adv}
\centering
\renewcommand{\arraystretch}{1.2}
\setlength{\tabcolsep}{3pt} 
\resizebox{\linewidth}{!}{ 
\small
\begin{tabular}{@{}c|ccc|ccc|ccc@{}}
\toprule
\multirow{2}{*}{\shortstack{Test \\ Setting}} 
& \multicolumn{3}{c|}{WER $\downarrow$} 
& \multicolumn{3}{c|}{Emo. Acc. (\%) $\uparrow$} 
& \multicolumn{3}{c}{Spk.-Sim. $\uparrow$} \\  
\cmidrule(l){2-10} 
& \method{} & \shortstack{$-$Emo. \\ Adv.}  & \shortstack{$-$Spk. \\ Adv.} 
& \method{} & \shortstack{$-$Emo. \\ Adv.}  & \shortstack{$-$Spk. \\ Adv.} 
& \method{} & \shortstack{$-$Emo. \\ Adv.}  & \shortstack{$-$Spk. \\ Adv.}     \\ 
\midrule
SSST & $\mathbf{5.58}$ & $5.91$ & $5.64$ & $\mathbf{0.69}$ & $0.58$ & $0.67$ & $\mathbf{0.74}$ & $0.74$ & $0.74$ \\
SSDT & $\mathbf{4.94}$ & $5.02$ & $5.01$ & $\mathbf{0.71}$ & $0.47$ & $0.67$ & $0.73$ & $\mathbf{0.75}$ & $0.74$ \\
DSST & $\mathbf{5.55}$ & $5.97$ & $5.73$ & $\mathbf{0.58}$ & $0.48$ & $0.56$ & $\mathbf{0.74}$ & $0.74$ & $0.72$ \\
DSDT & $\mathbf{6.48}$ & $6.59$ & $6.51$ & $\mathbf{0.56}$ & $0.32$ & $0.51$ & $\mathbf{0.72}$ & $0.7$ & $0.71$ \\ 
\bottomrule
\end{tabular}
}
\end{table}
\subsection{Importance of the adversarial losses}
An attentive reader might enquire about the necessity of the two adversarial losses in training \method{}, we analyze their impact using t-SNE visualizations~\cite{van2008visualizing} of x-vectors~\cite{desplanques20_interspeech} and the corresponding speaker embeddings for two speakers from the ESD dataset (see Fig.~\ref{fig:ease_ablation}). The t-SNE plots for x-vectors reveal well-defined clusters that correspond to emotion categories without the emotion adversarial training. To address this, we introduce an emotion adversarial loss (Eq.\ref{eq:lossspkr}) with $\lambda_{adv}^{emo} > 0$, forcing the speaker embeddings to be emotion-disentangled. The effectiveness of this approach is confirmed by the speaker embedding t-SNE plots in Fig~\ref{fig:ease_ablation}. \par
Additionally, we empirically assess the speaker adversarial loss in the emotion classifier (denoted as $\lambda_{adv}^{spk}$ in Eq.\ref{eq:lossemo}). The results of these ablation experiments are presented in Table~\ref{tab:adv}.\\
\textbf{Key takeaways}: 1) The removal of the emotion adversarial loss from the training of the speaker embedding leads to a significant drop in the emotion accuracy metric across all four test settings. This confirms that x-vectors inherently entangle speaker and emotion information, making disentanglement crucial. 2) The removal of the speaker adversarial loss from the emotion classifier (in Eq.~\ref{eq:lossemo}) leads to a drop in the speaker similarity for the test settings DSST and DSDT. 3) Removal of either of the adversarial losses increases WER, indicating that speaker-emotion disentanglement improves the quality of the converted speech.

\begin{table}[t!]

\caption{The weighted F1-scores for IEMOCAP $4$ and $6$ class classification under different training conditions and with or without augmentation. Orig. refers to using all the training samples, while $100$ and $200$ means only $100$ or $200$ samples from each class are used for training the model. No Aug. refers to when no augmented data is used, whereas experiments using synthesized samples are mentioned as Aug.}
\centering
\label{tab:aug}
\resizebox{\linewidth}{!}{
\begin{tabular}{l|ccc|ccc}
        \toprule
        Training data & \multicolumn{3}{c|}{IEMOCAP-4} & \multicolumn{3}{c}{IEMOCAP-6} \\
        \cmidrule(lr){2-4} \cmidrule(lr){5-7}
        & Orig. & $200$ & $100$ & Orig. & $200$ & $100$ \\
        \midrule
        No Aug. & $65.85$ & $59.33$ & $49.76$ & $51.67$ & $43.13$ & $36.41$ \\
        Aug. & $\mathbf{67.13}$ & $\mathbf{61.09}$ & $\mathbf{52.61}$ & $\mathbf{52.91}$ & $\mathbf{46.06}$ & $\mathbf{40.02}$ \\
        \bottomrule
    \end{tabular}}
\end{table}
\subsection{Application in Speech Emotion Recognition}
The emotion style transfer capability of \method{} can be leveraged for data augmentation in Speech Emotion Recognition (SER). To evaluate this, we use the IEMOCAP dataset\cite{busso2008iemocap}. This dataset consists of utterances grouped into $5$ sessions and categorized as one of $10$ emotions - ``happy'', ``angry'', ``sad'', ``neutral'', ``frustrated'', ``excited'', ``fearful'', ``surprised'', ``disgusted'' or ``other''. Following prior works~\cite{dutta2022multimodal, dutta2023hcam,dutta2024leveraging}, we conduct both 4-class and 6-class emotion classification on this dataset, using utterances from session $5$ as the test split and the other sessions for training and validation. For the four-way classification, the emotion categories considered are ``angry'', ``sad'', ``happy'' (merged with ``excited'') and ``neutral'', while the first $6$ emotion categories are considered for the six-way classification setting. All evaluations are carried out in a speaker-independent way. For SER model training, we extract features using WavLM-base\cite{chen2022wavlm} and follow the SUPERB evaluation framework\cite{yang2021superb}.

For augmentation, we select $100$ neutral utterances and $50$ utterances per emotion class from the training split of IEMOCAP. Using these emotional utterances as reference speech signals, we generate a total of $5000$ converted utterances per emotion class with \method{}. Since the converted utterances involve unseen speakers and reference speech signals, we apply a quality filtering based on the Emo.Sim. score (Sec.~\ref{sec:metrics}), where we take the top $100$ utterances for each of the emotion classes based on this score. Thus, $300$ utterances are used to augment IEMOCAP-4 class setting, while $500$ utterances are used for the IEMOCAP-6 class setting.

 To further investigate the impact of data augmentation, we create smaller versions of the IEMOCAP dataset - $100$ and $200$ training utterances per emotion class. We report the weighted F1-scores on all the test settings for this dataset in Table~\ref{tab:aug}.\\
 
\textbf{Key takeaways}: 1) In the full IEMOCAP-4 dataset, augmentation improves performance by $3.69\%$ (relative). However, for IEMOCAP-4($100$) (only $100$ samples per class), the improvement is $5.67\%$, highlighting the stronger impact of augmentation in low-resource settings. A similar trend is seen for IEMOCAP-6, where improvement jumps from $2.57\%$ (full set) to $5.68\%$ (with only $100$ samples per class). 2) Even though IEMOCAP-6 has two unseen emotion classes (``excited'' and ``fear''), \method{} improves for all the three test settings, indicating the generalizable nature of \method{}. 

\section{Summary}
\textit{Key Highlights:} This paper proposes an audio-to-audio emotion style transfer by introducing \method{}, an auto-encoding framework that reconstructs speech from emotional and non-emotional factors.  While pre-trained models are used for the content and speaker, the models are trained for the emotion embedding extraction. Additionally, we develop a pitch and duration prediction pipeline conditioned on these factors. These are combined by a speech synthesis module based on the BigVGAN architecture for speech reconstruction. During style transfer, content and speaker embeddings are derived from the source, whereas emotion embeddings—derived from a reference utterance—govern the pitch and duration of the converted speech. We evaluate \method{} through objective and subjective experiments under matched and mismatched speaker/text scenarios and demonstrate its generalization to unseen speakers and target emotions. \\
\textcolor{black}{\textit{Limitations and future scope:} The base \method{} configuration is trained on the English partition of the ESD dataset (approximately $15$ hours from $10$ speakers), which limits speaker diversity during training and affects generalization to unseen voices. While the \method{}-diverse variant demonstrates that increasing training data diversity improves speaker similarity and content preservation, performance in unseen-speaker scenarios remains below large-scale systems trained on substantially larger corpora. Future work will explore training on larger and more diverse emotional speech datasets.}

\bibliographystyle{IEEEbib}
\bibliography{refs}

\end{document}